\documentclass[12pt]{article}
\usepackage{amsmath,amssymb,amsfonts,amscd,cancel,graphicx,hyperref,longtable,bm,cite}
\begin{document}

%%%%%%%%%%%%%%%%%%%%%%%%%%%%%%%%%%%%%%%%%%%%%%%%%%%%%%%%%%%%%%%%%%%%%
%                     Title Page                                    %
%%%%%%%%%%%%%%%%%%%%%%%%%%%%%%%%%%%%%%%%%%%%%%%%%%%%%%%%%%%%%%%%%%%%%

\thispagestyle{empty}
\renewcommand{\thefootnote}{\fnsymbol{footnote}}
\setcounter{footnote}{1}

\vspace*{-0.5cm}

\centerline{\Large\bf Unbroken Discrete Supersymmetry} 

\vspace*{18mm}

\centerline{\large\bf
Gerhart Seidl\footnote{E-mail: \texttt{gerhart.seidl@gmail.com}}
}

\vspace*{5mm}
\begin{center}
{\em D 91207 Lauf a.d. Pegnitz, Germany} 
\end{center}

\vspace*{20mm}

\centerline{\bf Abstract}
\vspace*{2mm}
We demonstrate, by giving a specific example, that supersymmetry can be left unbroken without running into conflict with observation. The key idea is to employ a discrete form of supersymmetry. Amongst other interesting features, this construction goes roughly half the way in removing the hierarchy between the observed cosmological constant and the vacuum energies expected from field theory.
\renewcommand{\thefootnote}{\arabic{footnote}}
\setcounter{footnote}{0}

\newpage

%%%%%%%%%%%%%%%%%%%%%%%%%%%%%%%%%%%%%%%%%%%%%%%%%%%%%%%%%%%%%%%%%%%%%
%                     Main Text                                     %
%%%%%%%%%%%%%%%%%%%%%%%%%%%%%%%%%%%%%%%%%%%%%%%%%%%%%%%%%%%%%%%%%%%%%

\section{Introduction}

Paradoxically, the standard model provides too much and too little matter at the same time. Too little, because it only accounts for about 5\% of the matter-energy in the universe \cite{Ade:2013ktc}. Too much, because already the standard model vacuum energy overestimates the actually observed cosmological constant by about 120 orders of magnitude \cite{Riess:1998cb,Perlmutter:1998np,Spergel:2003cb}, which is also known as the cosmological constant problem \cite{Weinberg:1988cp,Carroll:2000fy,Nobbenhuis:2004wn}.

This paper presents a simple idea how to resolve this situation and kill two birds with one shot using discrete supersymmetry. How can we have more matter with less vacuum energy? Roughly, this can be answered by analogy with the supersymmetric harmonic oscillator \cite{Nicolai:1976xp,Witten:1981nf,Gendenshtein:1986ub}: Introduce supersymmetric partners that provide dark matter, but which cancel, by virtue of being superpartners, the vacuum energy. In this note, we will see that this remarkably simple idea goes a long way for usual matter with standard-model-type interactions.

In order to make the superpartners dark, we assume that they couple only via gravity to visible matter. This is possible for discrete supersymmetry and allows, at the same time, to leave it unbroken without implying sizable couplings to the superpartner fields. To not obscure the central mechanism of the model by the detailed structure of the full standard model, we illustrate all constituents, for simplicity, through a massive Majorana fermion that couples via a Yukawa interaction.

\section{Free Theory}\label{sec:discretesupersymmetry}
In this section, we will introduce a simple Fermi-Bose system with a discrete symmetry transformation interchanging fermionic with the bosonic degrees of freedom. We will show that this is an on- and off-shell symmetry of the total Lagrangian, with the kinetic and mass terms transforming independently, if and only if the bosons and fermions have the same mass. We will see that this mass degeneracy is stable under one-loop quantum corrections in Sec.~\ref{sec:loopcorrections}. Even though we will first concentrate on the free theory, later, in Sec.~\ref{sec:Yukawa}, we will extend the model by including a Yukawa interaction.

As announced, we illustrate the essential mechanism using a massive four-component Majorana field $\Psi^i\in{\bf C}^4$, $i=1,2,3,4$, plus superpartner fields consisting of two real physical scalar fields $A$ and $B$ and two auxiliary real scalar fields $F$ and $G$. As dynamics, we employ the four-dimensional free Wess-Zumino model \cite{Wess:1974tw}, which has the Lagrangian densities
\begin{eqnarray}\label{eq:scalarlagrangian1}
\mathcal{L}_\text{s}&=&\frac{1}{2}(\partial_\mu A\partial^\mu A+\partial_\mu B\partial^\mu B+F^2+G^2)\nonumber\\
&+&m(AF+BG)
\end{eqnarray}
and 
\begin{equation}\label{eq:fermionlagrangian}
\mathcal{L}_\text{f}=\frac{\text i}{2}\gamma^{\mu i}_{~~j}\overline{\Psi}_i\wedge\partial^\mu\Psi^j-\frac{m}{2}\overline{\Psi}_i\wedge \Psi^i,
\end{equation}
where $m$ is the mass, $\overline{\Psi}=\Psi^\dagger\gamma^0$, and $\gamma^\mu$, with $\mu=0,1,2,3$, are the gamma matrices. In (\ref{eq:fermionlagrangian}), the symbol $\wedge$ denotes the product for the Grassmann algebra over the four-dimensional complex vector space ${\bf C}^4$.  For the gamma matrices, we adopt the chiral representation, which is given by
\begin{equation}
\gamma^\mu=\left(
\begin{matrix}
0 & \sigma^\mu\\
\overline{\sigma}^\mu& 0
\end{matrix}
\right),\label{eq:chiral}
\end{equation}
where $\sigma^\mu=({\bf 1}_2,\bm{\sigma})^{\mu}$, $\overline{\sigma}^\mu=({\bf 1}_2,-\bm{\sigma})^{\mu}$, and $\sigma^i$ ($i=1,2,3$) are the Pauli matrices.

Let us now identify a discrete symmetry that maps $\mathcal{L}_\text{s}$ onto $\mathcal{L}_\text{f}$ and vice versa. For this purpose, we first organize the $c$-number-valued fields in (\ref{eq:scalarlagrangian1}) into a scalar four-component object $\xi^i\in{\bf R}^4$, $i=1,2,3,4$, as
\begin{equation}\label{eq:xi}
\xi=(A,B,F,G)^T,
\end{equation}
such that the scalar Lagrangian density in (\ref{eq:scalarlagrangian1}) becomes
\begin{equation}\label{eq:scalarlagrangian2}
\mathcal{L}_\text{s}=\frac{1}{2}\xi^T\left(
\begin{matrix}
-\partial_\mu\partial^\mu\cdot{\bf 1}_2&m\cdot{\bf 1}_2\\
m\cdot{\bf 1}_2 & {\bf 1}_2
\end{matrix}
\right)\xi.
\end{equation}
In addition, we will go to momentum space by using the Fourier coefficients $\tilde{\Psi}({p})$ and $\tilde{\xi}({p})$ of the fields, which are respectively defined by $\Psi(x)=\int[d^4p/(2\pi)^4]\tilde{\Psi}({p})e^{-{\text i}px}$ and $\xi(x)=\int[d^4p/(2\pi)^4]\tilde{\xi}({p})e^{-{\text i}px}$, where $x=(x^\mu)=(t,{\bf x})$ denotes the spacetime coordinates.

On the symmetric and the Grassmann algebra for the bosonic and fermionic fields, we can then define a discrete supersymmetry transformation $U$ that leaves the total free Lagrangian $\mathcal{L}_0=\mathcal{L}_\text{s}+\mathcal{L}_\text{f}$ invariant by mapping
\begin{equation}\label{eq:U}
U\,:\,\varphi_p(i,j) K(\tilde{\Psi}({p}))_i^\ast\wedge K(\tilde{\Psi}({p}))^j\leftrightarrow N(\tilde{\xi}({p}))^\ast_iN(\tilde{\xi}({p}))^j,
\end{equation}
where $K,N:\mathbf{C}^4\rightarrow{\mathbf{C}}^4$ are invertible linear transformations, $\varphi_p(i,j)$ is a momentum-dependent phase, and the indices $i$ and $j$ satisfy $i\leq j$, with $(i,j)$ taking the values $(i,i)$ or $(i,i+2)$. The cases $i>j$ follow after anticommuting the fermionic factors on the left-hand side. In (\ref{eq:U}), the value of the phase is $\varphi_p(i,j)=-1$ for $i\neq j$ and will be specified for $i=j$ as a function of $p$ below. Since the discrete symmetry in (\ref{eq:U}) exchanges products of fermionic and bosonic fields, we will call the mapping $U$ a discrete supersymmetry transformation.

Up to phases, the individual factors  in the product in (\ref{eq:U}) transform as
\begin{equation}\label{eq:Usingle}
U\,:\, K(\tilde{\Psi}({p}))^i\leftrightarrow N(\tilde{\xi}({p}))^i,
\end{equation}
and correspondingly for the complex conjugated fields. Since, however, the left-hand sides in (\ref{eq:U}) and (\ref{eq:Usingle}) belong to different exterior powers of the Grassmann algebra, we will employ (\ref{eq:U}) for a description of the discrete supersymmetry of the free theory. 

Representing in momentum space $K$ and $N$ respectively by invertible, momentum-dependent $4\times 4$ matrices $(K_{~j}^i)$ and $(N_{~j}^i)$, we assume, in the basis defined by (\ref{eq:chiral}) and (\ref{eq:xi}), that these satisfy for arbitrary momenta
\begin{subequations}\label{eq:squareroots}
\begin{eqnarray}
\varphi_p(i,j)(K^{\ast i}_{k} D_{~l}^k K_{~j}^l)&=&\text{diag}\left(\overline{\sigma}^\mu p_\mu,\sigma^\mu p_\mu\right),\label{eq:fermionroot}\\
(N^{\ast i}_k D_{~l}^k N_{~j}^l)&=&
\text{diag}\left(p^2\cdot{\bf 1}_2,{\bf 1}_2\right),\label{eq:scalarroot}
\end{eqnarray}
\end{subequations}
where $(D_{~l}^k)$ is the $4\times4$ matrix representation of a map $D:\mathbf{C}^4\rightarrow\mathbf{C}^4$ to be specified below, and $p_\mu$ is the momentum. We see from (\ref{eq:fermionroot}) that the role of $(K_{~j}^i)$ and $(N_{~j}^i)$ is to bring the fermionic and scalar kinetic terms to a common form, $(D_{~j}^i)$, which will turn out to be diagonal. We also observe in (\ref{eq:fermionroot}) that $(K_{~i}^j)$ appears quadratically in a product giving a first-order differential operator, and can, thus, be interpreted as a fractional derivative \cite{Oldham:1974}. 

\begin{figure}[t]
\centering
\includegraphics[width=0.30\textwidth]{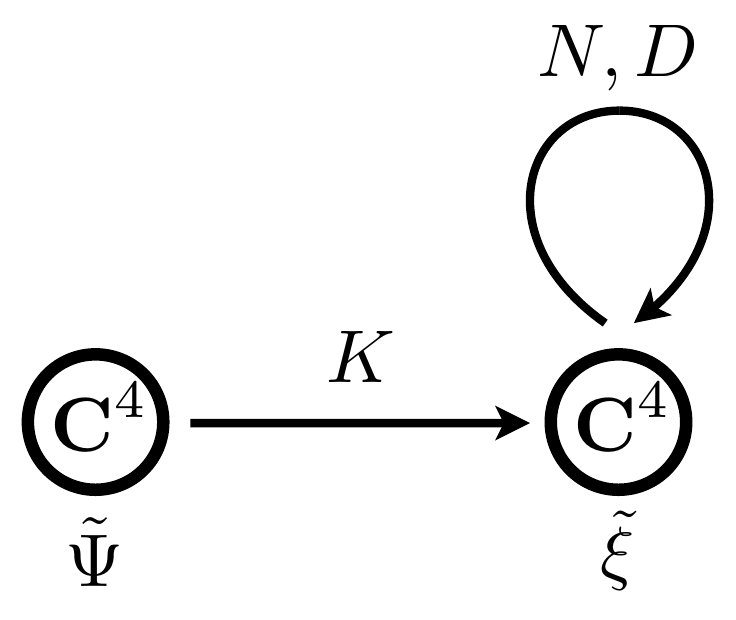}
\caption{Quiver representation for the mappings $K,N,$ and $D$ in momentum space. 
}\label{fig:Quiver}
\end{figure}

Fig.~\ref{fig:Quiver} shows a quiver representation (see, for example, \cite{Derksen:2005}) for $K,N,$ and $D$. The vertices correspond to the vector spaces for $\tilde{\Psi}({p})$ and $\tilde{\xi}({p})$, whereas the arrows represent the mappings $K,N,$ and $D$ between them. A closed loop denotes an endomorphism. Since both complex vector spaces are four-dimensional, this quiver representation has a dimension vector $(4,4)$. The figure illustrates that, as we will see in Sec.~\ref{sec:quantummechanics}, the mapping $K$ sends the vector space of spinor representations, which contains $\Psi$, onto a vector space of Lorentz singlets, where $\xi$ lives. The mappings $N$ and $D$ act on the latter vector space.

Applying for momentum $p$ the transformation in (\ref{eq:U}) and (\ref{eq:squareroots}) to the kinetic terms in (\ref{eq:fermionlagrangian}) and (\ref{eq:scalarlagrangian2}), we see that
\begin{eqnarray}\label{eq:kineticterms}
&&\hspace*{-8mm}\frac{1}{2}(\gamma^0\gamma^\mu p_\mu)^i_{~j}\tilde{\Psi}^\ast_i\wedge\tilde{\Psi}^j= 
\frac{1}{2}\sum_{i,j}\varphi_p(i,j)D_{~l}^k  K^{\ast i}_{k}\tilde{\Psi}^\ast_i\wedge K_{~j}^l\tilde{\Psi}^j\nonumber\\
&\leftrightarrow& \frac{1}{2}D_{~l}^k N^{\ast i}_{k}\tilde{\xi}^\ast_iN_{~j}^l\tilde{\xi}^j=\frac{1}{2}\tilde{\xi}^\dagger\left(
\begin{matrix}
p^2\cdot{\bf 1}_2&0\\
0&{\bf 1}_2
\end{matrix}
\right)\tilde{\xi},
\end{eqnarray}
where we have used that $(D^k_{~l})$ is diagonal, which leads to the case $i=j$ in (\ref{eq:U}), and suppressed the common argument $p$ of the Fourier coefficients. We thus find that $U$ maps the fermionic onto the scalar kinetic term and vice versa. Later, we will show that the mass terms are invariant under discrete supersymmetry, too.

Let us now give an explicit representation of $K,N,$ and $D$ in momentum space. These mappings have to satisfy the relations in (\ref{eq:squareroots}), which can be fulfilled for arbitrary on- and off-shell momenta $p$. In formulating these representations, we will distinguish the cases of timelike, lightlike, and spacelike momenta. When suitable, we will from now on denote the matrix representations by the same symbol as the corresponding mappings.

For timelike off-shell momenta $p$, we can choose $D={\bf 1}_4$ and
\begin{equation}
K=(p^2)^\frac{1}{4} S(\Lambda_p)^{-1},\,
N=\left(
\begin{matrix}
\sqrt{p^2}\cdot{\bf 1}_2&0\\
0&{\bf 1}_2
\end{matrix}
\right),\label{eq:timelike}
\end{equation}
where $S(\Lambda_p)$ is a $4\times 4$ spinor representation of the Lorentz transformation  $\Lambda_p:\Psi\rightarrow S(\Lambda_p)\Psi$ that takes the fermion from momentum $(p_0',0,0,0)^T$ in the rest frame to momentum $p$. 

If $p$ is timelike and on shell, we can take $D={\bf 1}_4$ and
\begin{equation}
K=(p^2)^\frac{1}{4}\mathcal{O}_\pm S(\Lambda_p)^{-1},\,
N=\mathcal{O}_-\left(
\begin{matrix}
\sqrt{p^2}\cdot{\bf 1}_2&0\\
0&{\bf 1}_2
\end{matrix}
\right),\label{eq:timelikeonshell}
\end{equation}
where $\mathcal{O}_\pm$ are the matrices
\begin{equation}
\mathcal{O}_\pm=\frac{1}{\sqrt{2}}\left(
\begin{matrix}
{\bf 1}_2 & \pm{\bf 1}_2\\
\mp{\bf 1}_2 & {\bf 1}_2
\end{matrix}
\right)\label{eq:O}
\end{equation}
In the first equation in (\ref{eq:timelike}), we assume $\mathcal{O}_+$ for $p_0>0$ and $\mathcal{O}_-$ for $p_0<0$. Actually, we require in (\ref{eq:O}) that the matrices $\mathcal{O}_\pm$  are only defined up to unitary rotations of the $2\times 2$ subspaces that are implied by the block structure. The matrices $\mathcal{O}_\pm$ serve the purpose to extract the physical components of $\tilde{\Psi}$ and $\tilde{\xi}$, for on-shell momenta. For example, $N(\tilde{\xi}({p}))$ becomes $N(\tilde{\xi}({p}))=\sqrt{2}m\big(\tilde{A}({p}),\tilde{B}({p}),0,0\big)^T$, if $p$ is on shell.

For spacelike momenta $p$, we can adopt $D=\text{diag}(-1,1,1,-1)$ and
\begin{equation}
K=|p^2|^\frac{1}{4}\Sigma S(\Lambda_p)^{-1},\; N=P_{13}\left(\begin{matrix}
\sqrt{|p^2|}\cdot{\bf 1}_2&0\\
0&{\bf 1}_2
\end{matrix}
\right),\label{eq:spacelike}
\end{equation}
where $S(\Lambda_p)$ is a Lorentz transformation matrix that sends the fermion from momentum $(0,p_0',0,0)$, with $|p_0'|=|\sqrt{-p^2}|$, to $p$ and $\Sigma=\frac{1}{\sqrt{2}}\text{diag}({\bf 1}_2-i\sigma^2,{\bf 1}_2-i\sigma^2)$, which is unitary and commutes with $\gamma^0$. Moreover, $P_{13}$ is the $4\times 4$ permutation matrix that interchanges the 1st with the 3rd row. The function of $\Sigma$ is to diagonalize $\overline{\sigma}^\mu p_\mu$ and $\sigma^\mu p_\mu$ in (\ref{eq:fermionroot}), and $P_{13}$ changes the ordering of the elements in $D$ in order to match the structure in (\ref{eq:scalarroot}).

Correspondingly, for lightlike momenta, we can pick $D=\text{diag}(0,1,1,0)$ and
\begin{equation}
K=\sqrt{2p_0}\Sigma^{-1} S(\Lambda_p)^{-1},\; N=P_{24}\left(
\begin{matrix}
2p_0\cdot{\bf 1}_2&0\\
0&{\bf 1}_2
\end{matrix}
\right),\label{eq:lightlike}
\end{equation}
where $S(\Lambda_p)$ denotes a Lorentz transformation matrix that takes the fermion from momentum $(p'_0,p'_0,0,0)$  to $p$ and $P_{24}$ is the $4\times 4$ permutation matrix that interchanges the 2nd with the 4th row.

For all the above timelike (on- and off-shell), spacelike, and lightlike cases, the phase $\varphi_p(i,i)$ in (\ref{eq:U}) and (\ref{eq:fermionroot}) becomes equal to $\varphi_p(i,i)=+1$, if $p'_0>0$, and $\varphi_p(i,i)=-1$, if $p'_0<0$.

Now, the mass terms get mapped under the supersymmetry transformation as
\begin{eqnarray}
&&\hspace*{-8mm}-\frac{m}{2}\gamma^{0i}_{~~j}\tilde{\Psi}^\ast_i\wedge\tilde{\Psi}^j=\frac{-m}{2\sqrt{|p^2|}}\overline{\gamma}^{0k}_{~~l}
{K^{~i}_{k}}^\ast\tilde{\Psi}_i^\ast\wedge K_{~j}^l\tilde{\Psi}^j\label{eq:massterms}\\
&\leftrightarrow&\frac{m}{2\sqrt{|p^2|}}\overline{\gamma}^{0k}_{~~l}N^{~i}_{k}\tilde{\xi}_i^\ast N_{~j}^l\tilde{\xi}^j
=\frac{1}{2}\tilde{\xi}^\dagger\left(
\begin{matrix}
0&m\cdot{\bf 1}_2\\
m\cdot{\bf 1}_2&0
\end{matrix}
\right)\tilde{\xi},\nonumber
\end{eqnarray}
where we have employed (\ref{eq:U}) for momentum $p$, defined $\overline{\gamma}^0=\mathcal{O}_\pm\gamma^0\mathcal{O}^T_\pm$, for $p$ on shell, and $\overline{\gamma}^0=\gamma^0$, otherwise, and used in the first equation in (\ref{eq:massterms}) that $K^\dagger \gamma_0 K/\sqrt{|p^2|}=\gamma_0$. For off-shell momenta, (\ref{eq:massterms}) is verified by recognizing that in this case $\varphi_p(i,j)=-1$. If $p$ is on-shell, we corroborate (\ref{eq:massterms}) by taking into account that $\overline{\gamma}^0$ is diagonal and undergoes a sign flip always along with $\varphi_p(i,i)$, which ensures consistency with the transformation of the kinetic terms in (\ref{eq:kineticterms}).

For arbitrary momenta, we therefore see that the supersymmetry transformation $U$ maps in (\ref{eq:fermionlagrangian}) and (\ref{eq:scalarlagrangian2}) the fermionic onto the scalar mass terms and the other way round. In addition, comparison with (\ref{eq:kineticterms}) shows that the kinetic and the mass terms transform independently under discrete supersymmetry. Observe also that the invariance of the total Lagrangian under $U$ is achieved without using the equations of motion. We therefore identify the discrete supersymmetry transformation with an off-shell symmetry \cite{Gates:1983nr}. Incidentally, the transformation is independent from any parameter of the theory, such as the mass, which is characteristic for off-shell symmetries \cite{Gates:1983nr,Sohnius:1985qm}. In Sec.~\ref{sec:Yukawa}, we will see that this holds in the interacting theory, too.

So far, discrete supersymmetry has been considered at the classical level, only. In the next section, we will discuss its implications for the quantum theory.

\section{Quantum Fields}\label{sec:quantummechanics}
In this section, we will analyze how discrete supersymmetry manifests itself at the quantum mechanical level and see that it provides a unitary Fermi-Bose symmetry.

We begin by using the equations of motion to eliminate the auxiliary fields $F$ and $G$ via $F=-mA$ and $G=-mB$ and consider for $A,B$, and the Majorana spinor $\Psi$ the Fourier expansions
\begin{subequations}\label{eq:fieldexpansions}
\begin{eqnarray}
A(x)&=&\int\frac{d^3 p}{(2\pi)^3 2 p_0} [a(p)e^{-\text{i}px}+a^\dagger(p) e^{\text{i}px}],\label{eq:quantizedA}\\
B(x)&=&\int\frac{d^3 p}{(2\pi)^3 2 p_0} [b(p)e^{-\text{i}px}+b^\dagger(p) e^{\text{i}px}],\label{eq:quantizedB}\\
\Psi(x)&=&\int\frac{d^3 p}{(2\pi)^3 2 p_0}\sqrt{m}\label{eq:quantizedfermion}\\
&&\hspace*{-12mm}\times\sum_{s=\pm\frac{1}{2}}S(\Lambda_p)[f_s({p})
\left(\begin{matrix}
\chi_s\\
\chi_s
\end{matrix}\right)e^{-\text{i}px}+f^\dagger_s({p})\left(\begin{matrix}
-\overline{\chi}_s\label{eq:quantizedPsi}\\
\overline{\chi}_s
\end{matrix}\right) e^{\text{i}px}],\nonumber
\end{eqnarray}
\end{subequations}
where $a$ and $b$ are the annihilation and $a^\dagger$ and $b^\dagger$ the creation operators of bosons that satisfy the usual bosonic commutation relations, $f_s$ and $f_s^\dagger$ are the anticommuting fermionic annihilation and creation operators , $p_0=\sqrt{m^2+\vec{p}^2}$, $\chi_{\pm\frac{1}{2}}\in {\bf C}^2$, with $\chi_i^\dagger\chi_j=\delta_{ij}$, and $\overline{\chi}_s=\text{i}\sigma^2\chi^\ast_s$ (for simplicity, we have set in (\ref{eq:quantizedfermion}) the creation phase factor equal to one).

Inserting these expansions into the Lagrangian densities in (\ref{eq:scalarlagrangian1}) and (\ref{eq:fermionlagrangian}), we find that the Lagrangians for the scalar and fermionic kinetic terms $L_\text{s}^\text{kin}$ and $L_\text{f}^\text{kin}$ become in the usual notation
\begin{subequations}\label{eq:Lagrangians}
\begin{eqnarray}
 L_\text{s}^\text{kin}&=&m^2\int\frac{d^3 p}{(2\pi)^2 p_0}[a({p}) a^\dagger({p})+b({p}) b^\dagger({p})],\\
 L_\text{f}^\text{kin}&=&-m^2\int\frac{d^3 p}{(2\pi)^2 p_0}\sum_{s=\pm\frac{1}{2}} f_s({p}) f^\dagger_s({p}),
\end{eqnarray}
whereas  the Lagrangians for the scalar and fermionic mass terms $L_\text{s}^\text{m}$ and $L_\text{f}^\text{m}$ take the forms
\begin{equation}
L_\text{s}^\text{m}=-L_\text{s}^\text{kin},\quad L_\text{f}^\text{m}=-L_\text{f}^\text{kin}.
\end{equation}
\end{subequations}
In (\ref{eq:Lagrangians}), we have omitted constant commutator terms, which cancel between bosons and fermions, anyway.

Now, we can read off from (\ref{eq:U}) that the discrete supersymmetry transformation amounts for the expansions in (\ref{eq:fieldexpansions}) to a untiary permutation symmetry that is given by
\begin{equation}\label{eq:operatorproducts}
f_1^\dagger({p})f_1({p})\leftrightarrow a^\dagger({p})a({p}),\quad f^\dagger_2({p}) f_2({p})\leftrightarrow b({p})^\dagger b({p}).
\end{equation}
Similarly, we find from (\ref{eq:quantizedfermion}) that (\ref{eq:Usingle}) translates into
\begin{eqnarray}\label{eq:singleoperators}
f_1({p})\leftrightarrow a({p}),&&f_2({p})\leftrightarrow b({p}),\nonumber\\
f_1^{\dagger}({p})\leftrightarrow a^{\dagger}({p}),&&f_2^{\dagger}({p})\leftrightarrow b^{\dagger}({p})
\end{eqnarray}
where we have for the annihilation operators made use of the freedom to redefine the matrix $\mathcal{O}_-$ in the expression for $K$ in (\ref{eq:timelikeonshell}) as $\mathcal{O}_-\rightarrow V\mathcal{O}_-$, with the block-diagonal unitary matrix $V=\text{diag}(\text{i}\sigma^2,\text{i}\sigma^2)$, which leaves all previous results unchanged.

After anticommuting the fermionic operators in (\ref{eq:operatorproducts}), we see that the Lagrangians in (\ref{eq:Lagrangians}) are indeed invariant under the permutation symmetry in (\ref{eq:operatorproducts}). Discrete supersymmetry therefore amounts at the quantum level to a unitary symmetry of the free theory in agreement with Wigner's theorem \cite{Wigner:1931}.

Inasmuch as discrete supersymmetry links different representations of the Lorentz group, it is instructive to analyze its relation to Lorentz invariance. In doing so, we will follow the example given in \cite{Peskin:1995ev} (see also \cite{Weinberg:1995mt}). For this purpose, let us denote by $W(\Lambda)$ the unitary operator that realizes a Lorentz transformation $\Lambda$ on the Hilbert space of physical one-particle states $|{\bf p},s\rangle=\sqrt{2E_p}f_s^\dagger({p})|0\rangle$, where $E_p=p_0$. Recall that Lorentz-invariance of the inner product $\langle{\bf p},r|{\bf q},s\rangle$ then implies that $f^{(\dagger)}_s({p})$ transforms under $\Lambda$ as $f^{(\dagger)}_s({p})\rightarrow W(\Lambda)f^{(\dagger)}_s({p})W^{-1}(\Lambda)=[E_{\Lambda p}/E_p]^\frac{1}{2}f^{(\dagger)}_s({p})$, where it has been assumed that the axis of spin quantization is parallel to the boost or rotation axis \cite{Peskin:1995ev}. Next, consider $K(\Psi(x))$, which is defined Fourier-component-wise via (\ref{eq:Usingle}). Since the matrix $K$ eliminates in (\ref{eq:quantizedfermion}) the spinor transformation matrix $S(\Lambda_p)$ from each Fourier coefficient $\tilde{\Psi}({p})$, we find that
\begin{equation}
W(\Lambda)K(\Psi(x))W^{-1}(\Lambda)=K(\Psi(\Lambda x)),
\end{equation}
that is, $K(\Psi(x))$ transforms (passively) as a scalar under the Lorentz group. As illustrated by the quiver representation in Fig.~\ref{fig:Quiver}, the mapping $K$ can therefore be viewed as sending the Lorentz spinor $\Psi(x)$ onto a subspace of ${\bf C}^4$ that contains Lorentz-singlets. This also implies that the Lorentz group is an isomorphism of the quiver representation.

Correspondingly, after integrating out the auxiliary fields, it follows from (\ref{eq:quantizedA}) and (\ref{eq:quantizedB}) that $R(\xi)$, where $R\equiv K^{-1}\circ N$, transforms as a spinor:
\begin{equation}
W(\Lambda)R(\xi(x))W^{-1}(\Lambda)=S^{-1}(\Lambda)R(\xi(\Lambda x)).\label{eq:spinortransformation}
\end{equation}
Discrete supersymmetry maps therefore Lorentz-invariant operators constructed from $\xi$ and $\Psi$ onto operators that are again Lorentz-invariant.

Until now, we have discussed discrete supersymmetry for the free theory, only. In the next section, we will show how it carries over to the interacting theory for the case of a renormalizable bosonic interaction between the fermions.

\section{Interactions}\label{sec:Yukawa}

In this section, we will include an interaction for the discrete supersymmetry model. For simplicity, we will first confine ourselves to a Yukawa interaction, which serves as a simple representative for the known interactions mediated by bosons, such as the Higgs and the gauge bosons in the standard model. In the following, we will work in the interaction picture (see, however, \cite{Haag:1955ev}).

To implement a Yukawa interaction, we introduce a copy of the matter sector discussed in Sec.~\ref{sec:discretesupersymmetry}. This new sector is described by representations with scalar and fermionic Lagrangians $\mathcal{L}'_\text{s}$ and $\mathcal{L}'_\text{f}$ that are identical with those given in (\ref{eq:fermionlagrangian}) and (\ref{eq:scalarlagrangian2}), but where all fields are now replaced by new, primed fields $A',B'$ (along with the corresponding auxiliary fields $F'$ and $G'$) and $\Psi'$, which have a primed mass $m'$, and so on. This sector is then invariant under the discrete supersymmetry transformation (\ref{eq:U}), formulated for the primed fields.

We let the primed sector couple to the fermion $\Psi$, introduced in Sec.~\ref{sec:discretesupersymmetry}, via the Yukawa interaction
\begin{equation}\label{eq:Yukawa}
\mathcal{L}_\text{Y}=\frac{1}{2}\eta_l^\ast{\xi'}^l\gamma^{0i}_{~~j}\Psi^\ast_i\wedge\Psi^j
+\text{h.c.},
\end{equation}
where
\begin{equation}\label{eq:coupling}
(\eta_l)=(m'\eta_a,m'\eta_b,\eta_c,\eta_d),
\end{equation}
for complex $\eta_a, \eta_b,\eta_c,$ and $\eta_d$. In (\ref{eq:Yukawa}), the scalar field $\xi'$ is defined in complete analogy with (\ref{eq:xi}).

Applying the equations of motion $F'=-m'A'$ and $G'=-m'B'$ to eliminate the primed auxiliary fields, we see that (\ref{eq:Yukawa}) reduces to the usual form for a Yukawa interaction with two real scalar fields
\begin{equation}\label{eq:usualYukawa}
\mathcal{L}_\text{Y}=(Y_aA'+Y_bB')\gamma^{0i}_{~~j}\Psi^\ast_i\wedge\Psi^j,
\end{equation}
where $Y_a=\text{Re}(\eta_a-\eta_c)$ and $Y_b=\text{Re}(\eta_b-\eta_d)$. In momentum space, the Yukawa interaction takes the form
\begin{eqnarray}
\int d^4x\,\mathcal{L}_\text{Y}&=&\frac{1}{2}\int \frac{d^4k\,d^4p\,d^4q}{(2\pi)^8}\,\delta^{(4)}(k+p+q)\nonumber\\
&&\hspace*{-4.8mm}\times\,\eta_l^\ast \tilde{\xi}'(k)^l\gamma^{0i}_{~~j}\tilde{\Psi}({p})^\ast_i\wedge\tilde{\Psi}({q})^j+\text{h.c.}\label{eq:LYtilde}
\end{eqnarray}
From (\ref{eq:Usingle}), we obtain that $\mathcal{L}_\text{Y}$ is mapped by the discrete supersymmetry transformation onto a new interaction $\overline{\mathcal{L}}_\text{Y}$ for $\xi$ that is in momentum space given by
\begin{eqnarray}
\int d^4x\,\overline{\mathcal{L}}_\text{Y}&=&\frac{1}{2}\int \frac{d^4k\,d^4p\,d^4q}{(2\pi)^8}\,\delta^{(4)}(k+p+q)\label{eq:LYbartilde}\\
&&\hspace*{-4.8mm}\times\,\eta^\ast_lR^{-1}(\tilde{\Psi}'(k))^l\gamma^{0i}_{~~j}R(\tilde{\xi}({p}))^\ast_i R(\tilde{\xi}({q}))^j+\text{h.c.},\nonumber
\end{eqnarray}
which contains on shell only a single three-dimensional momentum conservation delta function and $R$ has been defined before (\ref{eq:spinortransformation}). The new interaction can be written in configuration space as
\begin{equation*}
\overline{\mathcal{L}}_\text{Y}(x)=\frac{1}{2}\eta^\ast_lR^{-1}(\Psi'(x))^l\gamma^{0i}_{~~j}R(\xi(x))^\ast_i R(\xi(x))^j+\text{h.c.},
\end{equation*}
where all fields are taken at the same spacetime point, $x$, and $R$ and $R^{-1}$ are interpreted as functions of derivatives. A subsequent supersymmetry transformation then leads from $\overline{\mathcal{L}}_\text{Y}$ back to the interaction $\mathcal{L}_\text{Y}$.

From our discussion in Sec.~\ref{sec:quantummechanics}, we see that the interaction $\overline{\mathcal{L}}_\text{Y}$ for the quantized fields is Lorentz-invariant. For the expansions in (\ref{eq:fieldexpansions}), it also follows that discrete supersymmetry provides a unitary symmetry between $\mathcal{L}_\text{Y}$ and $\overline{\mathcal{L}}_\text{Y}$. Moreover, by applying discrete supersymmetry to the Hamiltonian density $\mathcal{H}$ that corresponds to $\mathcal{L}_\text{f}+\mathcal{L}_\text{f}+\mathcal{L}_\text{Y}$, we find that the interaction Hamiltonian density of the supersymmetric partner fields, $\overline{\mathcal{H}}_\text{int}$, is, as usual, given by $\overline{\mathcal{H}}_\text{int}=-\overline{\mathcal{L}}_\text{Y}$, which can be used for a Dyson series expansion of the $S$-matrix (for a pedagogical discussion concerning the possible different types of interaction Hamiltonians, see \cite{Sakurai:2011}). Note also that since the operator $W(\Lambda)$ (see Sec.~\ref{sec:quantummechanics}) commmutes with the symmetry transformations in (\ref{eq:singleoperators}), we find from the $S$-matrix expansion (cf.~Sec.~\ref{sec:loopcorrections}) that if $W^\dagger(\Lambda) SW(\Lambda)=S$, where $S$ is the $S$-matrix for the Yukawa interaction Hamiltonian $\mathcal{H}_\text{int}$, the $S$-matrix for the new interaction $\overline{\mathcal{H}}_\text{int}$ will commute with Lorentz transformations, too.

Interaction Lagrangians that are qualitatively similar to $\overline{\mathcal{L}}_\text{Y}$ have been studied in \cite{Soroka:1995et}. Let us emphasize, nonetheless, that we are not restricted to this kind of tri-linear interactions. Instead, we could assume other types of interactions, such as in the Nambu-Jona-Lasinio model \cite{Nambu:1961tp}, which would predict from the symmetry new interaction Lagrangians that involve, for example, only scalar fields. To be definite, we shall nevertheless stick in the following mainly to the Yukawa interaction outlined above.

The total matter Lagrangian density of the interacting model, invariant under discrete supersymmetry, therefore reads
\begin{equation}
\mathcal{L}_\text{m}=\underbrace{\mathcal{L}_\text{f}+\mathcal{L}'_\text{s}+\mathcal{L}_\text{Y}}_\text{ordinary fields}+
\underbrace{\mathcal{L}_\text{s}+\mathcal{L}'_\text{f}+\overline{\mathcal{L}}_\text{Y}}_\text{superpartners},\nonumber
\end{equation}
where $\mathcal{L}_\text{f},\mathcal{L}'_\text{s},$ and $\mathcal{L}_\text{Y}$ define a visible sector of ordinary fields and $\mathcal{L}_\text{s},\mathcal{L}'_\text{f},$ and $\overline{\mathcal{L}}_\text{Y}$ a new, dark sector of superpartner fields. The subscript ``m'' stands for ``matter'', that is, the non-gravitational part of the Lagrangian.  In the standard model, the fermion $\Psi$ represents our usual matter fields, that is, the quarks and leptons, whereas their scalar superpartners, corresponding to $\xi$, are mass degenerate, but cannot be observed directly, for there are no direct interactions between the usual fields and the exotics. This is the central feature that allows discrete supersymmetry to be left unbroken, without running into immediate conflict with observation. In the standard model, for example, the photon would be replaced in the dark sector by a fermionic superpartner that interacts only with the dark sector. This implies, for example, that gauge coupling unification in the discrete supersymmetric theory would be just as in the standard model.

\begin{figure}[t]
\centering
\includegraphics[width=0.30\textwidth]{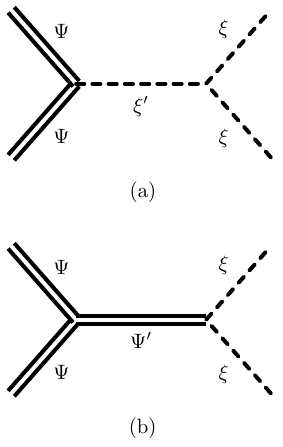}
\caption{Communication between ordinary and  their superpartner fields. Discrete supersymmetry predicts from the couplings in (a) the new interactions in (b).}\label{fig:Yukawa}
\end{figure}

In the minimal formulation, the new fields interact with the ordinary ones only via gravity with couplings that are suppressed by the Planck scale $M_\text{Pl}\simeq 10^{18}\,\text{GeV}$. It may therefore be possible to probe the new predicted particles, for example, for a lowered Planck scale \cite{ArkaniHamed:1998rs,Randall:1999ee}. Moreover, extra messenger fields could, similar to gravity, communicate between the ordinary and the new fields and lead to additional observable effects.

To establish an interaction between the usual and their superpartner fields in this way, let us assume that $\xi'$ couples to $\xi$ as
\begin{equation}
\overline{\mathcal{L}}_\lambda=-\lambda_l^\ast{\xi'}^l\gamma^{0i}_{~~j}\xi_i\xi^j+\text{h.c.},\label{eq:hiddencoupling}
\end{equation}
where $\lambda'$ is a 4-component object similar to $\eta'$. The vertices for $\overline{\mathcal{L}}_\lambda$ and $\mathcal{L}_\text{Y}$ are shown in Fig.~\ref{fig:Yukawa} (a). (We adopt the convention that a double line represents a Majorana fermion. Feynman rules for Majorana fermions are discussed, for example, in \cite{Haber:1984rc,Denner:1992vza,Mohapatra:1998rq,Dreiner:2008tw}.) Similarly to (\ref{eq:LYbartilde}), discrete supersymmetry then predicts from $\overline{\mathcal{L}}_\lambda$ a new interaction $ \mathcal{L}_\lambda$ among the ordinary fields that reads in momentum space
\begin{eqnarray}
 \int d^4x\,\mathcal{L}_\lambda&=&\int \frac{d^4k\,d^4p\,d^4q}{(2\pi)^8}\,\delta^{(4)}(k+p+q)\label{eq:newinteraction}\\
&&\hspace*{-2cm}\times\,\lambda^\ast_l R^{-1}(\tilde{\Psi}'(k))^l\gamma^{0i}_{~~j}R^{-1}(\tilde{\Psi}({p}))^\ast_i R^{-1}(\tilde{\Psi}({q}))^j+\text{h.c.},\nonumber
\end{eqnarray}
which is, again, seen to be Lorentz-invariant when inserting the field expansions in (\ref{eq:fieldexpansions}) and contains on shell only one single three-dimensional momentum conservation delta function. The vertices for $\mathcal{L}_\lambda$ and $\overline{\mathcal{L}}_\text{Y}$ are depicted in Fig.~\ref{fig:Yukawa} (b). Depending on the structure of $\overline{\mathcal{L}}_\lambda$, there are, of course, more general possibilities for the coupling between $\Psi$ and $\Psi'$ than in (\ref{eq:newinteraction}), such as vector- and axial-vector couplings, which could be phenomenologically interesting, but we will not further expand on this here. 

Observe that the coupling $\lambda$ can be naturally small in an extension of the model where, for instance, the visible sector contains two generations of Dirac fermions $\Psi_1$ and $\Psi_2$. If, for example, $\Psi_1,\Psi_2,$ and $\xi'$ carry the charges $+1,+2$, and $-1$ under some cyclic group $G=Z_n$ ($n\geq 3$), the fermions $\Psi_1$ and $\Psi_2$ will have nonzero Dirac masses and exhibit a renormalizable Yukawa interaction similar to (\ref{eq:Yukawa}). At the same time, the scalar superpartners for $\Psi_1$ and $\Psi_2$ can have charges that are different from the fermions if $G$ does not commute with discrete supersymmetry. In such a case, the coupling $\lambda$ can be forbidden by requiring, for example, that  the sfermions be singlets under $G$. A small non-vanishing $\lambda $ would then result if $G$ is only approximately conserved \cite{Froggatt:1978nt}.

Further interactions between the visible and the dark sector could come from quartic terms that mix visible scalar fields with dark sfermions. In the standard model, the only visible such scalar is the Higgs field. Theoretical uncertainties in the branching ratios of the Higgs (such as $H\rightarrow\gamma\gamma$) are of the order 5\%, which will be hard to accomplish at the LHC
 \cite{Beringer:1900zz}. A tuning of the quartic couplings roughly at the $1\%$-level (or a little less) should therefore be sufficient to escape experimental bounds in the near future. This degree of fine-tuning is, however, utterly insignificant as compared with the at least 60 orders of magnitude that discrete supersymmetry improves upon the cosmological fine-tuning-problem of the standard model (see Sec.~\ref{sec:darkenergy}) and can therefore be neglected safely.

One feature of the current model is that since quartic self-interactions are absent, the scalar superpartner sector would not suffer from the problem of quadratic divergencies, which destabilize the masses in usual scalar field theories. The behavior of the mass operators at the loop level in our model will be discussed in the next section.

\section{Loop Corrections}\label{sec:loopcorrections}
\begin{figure}[t]
\centering
\includegraphics[width=0.30\textwidth]{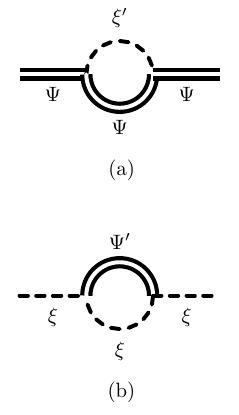}
\caption{Fermion (a) and sfermion (b) self-energies for the fields $\Psi$ and $\xi$ in the discrete supersymmetry model at one loop.}\label{fig:loops}
\end{figure}
We will now now analyze in how far the predictions of discrete supersymmetry are stable under quantum corrections by studying the running of the boson and fermion masses. As a consequence of the structure of the supersymmetric couplings, we will argue that fermions and their scalar superpartner fields exhibit the same running mass parameters at one loop, which means that the classical Fermi-Bose mass degeneracy described by discrete supersymmetry remains intact at the quantum level.

The one-loop corrections to the fermion and sfermion propagators in the discrete model are shown in Fig.~\ref{fig:loops}. While Fig.~\ref{fig:loops} (a) displays the usual fermion self-energy in standard Yukawa theory, Fig.~\ref{fig:loops} (b) represents the one-loop diagram arising in the dark sector. In the Wess-Zumino model, the quadratic divergence of the sfermion mass famously cancels the divergence coming from the sfermion mass correction involving a fermion loop. These quadratically divergent diagrams are absent in the minimal discrete supersymmetry model, which has no scalar quartic term or a sfermion-fermion-fermion coupling. The scalar sector is, therefore, free from the usual problem of quadratic divergencies. Instead, both the diagrams in Fig.~\ref{fig:loops} are only logarithmically divergent, just as the corresponding one-loop diagrams responsible for the running of the masses in the Wess-Zumino model.

\begin{figure}[t]
\centering
\includegraphics[width=0.25\textwidth]{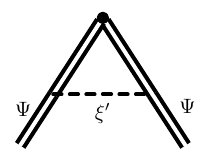}
\caption{One-loop correction to the fermion mass operator $\frac{1}{2}m\overline{\Psi}_i\wedge \Psi^i$. The dot denotes the fermion mass insertion.}\label{fig:massoperator}
\end{figure}

In the visible sector, we calculate the running of the fermion mass by treating the fermion mass operator as a small perturbation to the massless theory \cite{Peskin:1995ev2}. To this end, let us write $\mathcal{L}_\text{f}$ as
\begin{eqnarray}\label{eq:barelagrangian}
\mathcal{L}_\text{f}&=&\frac{\text i}{2}\gamma^{\mu i}_{~~j}\overline{\Psi}_i\wedge\partial_\mu\Psi^j-\frac{1}{2}m'\overline{\Psi}_i\wedge \Psi^i\nonumber\\
&+&\frac{\text i}{2}\delta_2\gamma^{\mu i}_{~~j}\overline{\Psi}_i\wedge\partial_\mu\Psi^j-\frac{1}{2}\delta_{m'}\overline{\Psi}_i\wedge \Psi^i\nonumber\\
&-&\frac{1}{2}m\overline{\Psi}_i\wedge \Psi^i-\frac{1}{2}m\delta_{\overline{\Psi}\Psi}\overline{\Psi}_i\wedge \Psi^i,
\end{eqnarray}
where $m'$ is the ``underlying mass'', which is renormalized to zero, $\delta_2$ is the counterterm for the fermion field strength renormalization, $\delta_{m'}$ is the mass counterterm for $m'$, and $\delta_{\overline{\Psi}\Psi}$ is the counterterm for the fermion mass operator $\frac{1}{2}m\overline{\Psi}_i\wedge \Psi^i$. From Fig.~\ref{fig:loops} (a) and the one-loop correction to the mass operator shown in Fig.~\ref{fig:massoperator}, we obtain in dimensional regularization
\begin{equation}\label{eq:counterterms}
\delta_2=-2\frac{\eta^T\eta}{(4\pi)^2}\frac{\Gamma(2-d/2)}{(\mu^2)^{2-d/2}},\quad\delta_{\overline{\Psi}\Psi}=-2\delta_2,
\end{equation}
where $\Gamma$ is the gamma function, $\mu$ the renormalization scale, and $d\rightarrow 4$. The counterterms $\delta_2$  and $\delta_{\overline{\Psi}\Psi}$ are, respectively, 4 and $-4$ times the corresponding result for a Dirac fermion with gauge interaction. As a result, the anomalous dimension $\gamma_{\overline{\Psi}\Psi}$ of the fermion mass operator is to one-loop order given by
\begin{equation}
\gamma_{\overline{\Psi}\Psi}=\mu\frac{\partial}{\partial \mu}(-\delta_{\overline{\Psi}\Psi}+\delta_2)=\frac{12}{(4\pi)^2}(Y_A^2+Y_B^2).\label{eq:anomalous}
\end{equation}
To leading order in the Yukawa couplings, the running (effective) fermion mass therefore reads
\begin{equation}\label{eq:runningmass}
\overline{m}(Q)=m\big{(}1+\gamma_{\overline{\Psi}\Psi}\text{ln}(Q/\mu)\big{)},
\end{equation}
where $Q$ is the momentum scale at which the effective mass is evaluated, $m=\overline{m}(\mu)$, and $\gamma_{\overline{\Psi}\Psi}$ is given in (\ref{eq:anomalous}).

Let us next consider the running of the scalar mass in the dark sector. The two-point Green's function in the dark sector is, as usual,
\begin{equation}\label{eq:G}
G(x,y)=\frac{\langle 0| T\{\xi(x)\xi(y)S\}|0\rangle}{\langle 0|S|0\rangle},
\end{equation}
where $T$ is the time-ordering operator and $S$ denotes the S-matrix in the dark sector, which exhibits a Dyson series for $\overline{H}_\text{Y}(t)=-\int d^3x\,\overline{\mathcal{L}}_\text{Y}(x)$ (cf.~Sec.~\ref{sec:Yukawa}) that can be written as
\begin{eqnarray}
S&=&\sum_{n=0}^\infty(-i)^n\int_{-\infty}^\infty dt_1\int_{-\infty}^{t_1}dt_2\dots\int_{-\infty}^{t_{n-1}}dt_n\nonumber\\
&&\times\,\overline{H}_\text{Y}(t_1)\overline{H}_\text{Y}(t_2)\dots\overline{H}_\text{Y}(t_n).\label{eq:Smatrix}
\end{eqnarray}
 From the free part of the theory it follows that discrete supersymmetry maps in momentum space the fermionic and scalar propagators  as
\begin{equation}\label{eq:propagators}
S_\text{F}({p})\rightarrow {R^{-1}}^\dagger S_\text{F}({p})R^{-1},\quad
\Delta'({p})\rightarrow R^\dagger \Delta'({p})R,
\end{equation}
where $S_\text{F}({p})$ and $\Delta'({p})$ are the propagators of $\Psi$ and $\xi'$, respectively, and $R=K^{-1}N$ has been introduced in (\ref{eq:spinortransformation}) and is evaluated at the momentum $p$. Simultaneously, the fermion mass terms get transformed as in (\ref{eq:massterms}), where $N$ can be equivalently replaced by $R$, and the Yukawa coupling as in (\ref{eq:LYbartilde}). The counterterms are mapped correspondingly under the discrete supersymmetry transformation.

After application of Wick's theorem and insertion of the above expressions into (\ref{eq:G}) in momentum space, it is seen that the matrices $R$ and $R^{-1}$ drop out of the loop integrals associated with Fig.~\ref{fig:loops} (b) and the diagram analogous to Fig.~\ref{fig:massoperator} for the mass correction of $\xi$. This is because the matrices $R$ and $R^{-1}$ that multiply the couplings in the dark sector get divided out by respective factors $R^{-1}$ and $R$ coming from the internal lines in (\ref{eq:propagators}) that are attached to them.

As a consequence, the Green's function (\ref{eq:G}) in the dark sector is reduced to the fermionic two-point function in the visible sector. The dependence of the scalar two-point function on $\eta$, $m$, and $\mu$ is, therefore, equal to that for the fermion. This implies that the fermion $\Psi$ and its scalar superpartner $\xi$ satisfy the same Callan-Symanzik equation and that they have, in particular, identical anomalous dimensions.

We thus infer that the scalar mass parameter $m$ in (\ref{eq:scalarlagrangian1}) and (\ref{eq:scalarlagrangian2}) will evolve exactly as the fermion mass. At one-loop, the running masses of the scalars $A$ and $B$ are, accordingly, degenerate with the fermion mass $\overline{m}(Q)$ given in (\ref{eq:runningmass}). As a consequence, when invariance of the total Lagrangian under discrete supersymmetry is imposed at some energy scale, the Fermi-Bose mass degeneracy remains intact under renormalization group running.

This result is, in fact, as expected: Since discrete supersymmetry has not been gauged, anomalies are absent, when ignoring wormhole effects \cite{Krauss:1988zc}. The classical symmetry ensuring the  Fermi-Bose mass degeneracy must therefore carry over to the quantum theory. Of course, it would be important to investigate further the relation of discrete supersymmetry to  the usual perturbative non-renormalization theorems \cite{Grisaru:1979wc}. Such a discussion, however, would require an extended analysis that has to be left for elsewhere.

So far, we have been focussing on the microscopic behavior of discrete supersymmetry. In the next section, we will analyze its impact on cosmology.

\section{Cosmological Implications}\label{sec:cosmology}
Next, we will briefly discuss the impact of discrete supersymmetry on cosmology. In this respect, our model may be relevant in at least two ways: dark matter \cite{Bernabei:2003za,Ahmed:2009zw,Angle:2007uj} (for an overview, see, for example, \cite{Trimble:1987ee,Hooper:2009zm}) and dark energy \cite{Riess:1998cb,Perlmutter:1998np,Spergel:2003cb}.

\subsection{Dark Matter}
First, the dark sector provides stable scalar or fermionic matter species with masses $m$ and $m'$. In an analogous construction for the SM, $m$ and $m'$ are then expected to be of the order $\lesssim1\,\text{GeV}$, which lies in the range of light dark matter (LDM) \cite{Boehm:2003hm}. Furthermore, this matter resides in a sector that interacts, in the minimal formulation, only super-weakly with the visible sector fields through gravity. The matter in the dark sector could, thus, provide super-WIMPs \cite{Pospelov:2008jk} as candidates for LDM, similar to scenarios of mirror dark matter \cite{Foot:2003iv}. As in models where the gravitino serves as dark matter, it may be necessary that the super-WIMPs result in our case from the decay of a next-to-lightest supersymmetry particle.

If we switch on an extra interaction between the visible and the dark sector as described in Sec.~\ref{sec:Yukawa}, the Lee-Weinberg bound \cite{Lee:1977ua} indicates that the mass $m'$ of the associated mediator particle should be lighter than the $Z$-boson mass. The interaction with the hidden sector would then correspond to a new, ``dark force'' \cite{Holdom:1985ag,ArkaniHamed:2008qn}.

Note that the observed difference in the baryonic and dark matter relic densities $\Omega_\text{B}\simeq 0.05$ and $\Omega_\text{DM}\simeq 0.25$ does not imply that discrete supersymmetry would be broken. This is because the Witten index \cite{Witten:1982df}, telling us whether supersymmetry is broken or not, counts only the difference between the zero fermionic and bosonic energy states and is independent from the actual occupation numbers of bosons and fermions.

The dark matter relic density could, in fact, be understood in a variant of spontaneous baryogenesis \cite{Cohen:1988kt} for the dark matter sector. In this scenario, a new inflaton-type scalar field $\phi$ couples to a dark matter current, thereby introducing a chemical potential for dark matter, when $\phi$ is initially displaced from its equilibrium point.  Different initial conditions for the fields and their supersymmetric partners  could then account for the distinct relic densities of ordinary and dark matter while leaving discrete supersymmetry unbroken.

\subsection{Dark Energy}\label{sec:darkenergy}
Ever since its inception, it has been known that the enhanced UV behavior of unbroken supersymmetry may be related to the cosmological constant problem \cite{Zumino:1974bg}. As a matter of fact, a bare cosmological constant receives contributions from the vacuum energy $\rho=\langle 0| T^{00}|0\rangle$, where $T^{00}$ denotes the 00-component of the stress-energy tensor
$T^{\mu\nu}=2 (\delta\mathcal{L}_\text{m}/\delta g_{\mu\nu})+g^{\mu\nu}\mathcal{L}_\text{m}$, with the metric $g_{\mu\nu}$, for the (non-gravitational) matter fields \cite{Zeldovich:1967gd}. Now, continuous supersymmetry \cite{Wess:1974tw,Ramond:1971gb,Golfand:1971iw,Volkov:1972jx,Nilles:1983ge,Haber:1984rc,Wess:1992cp,Martin:1997ns}, when unbroken, implies that the vacuum energy is zero.

Vanishing contributions to the vacuum energy are, however, not a hallmark of unbroken continuous supersymmetry only, but may also appear in discrete supersymmetry. In order to see this, let us repeat Pauli's argument \cite{Pauli:2000} (cf.~\cite{West:1986wua}): The vacuum energy for a particle with spin $j$ and mass $m_j$ is
\begin{eqnarray}
&&\hspace*{-3mm}\rho=\frac{1}{2}(-1)^{2j}(2j+1)\int d^3 k \sqrt{k^2 +m_j^2}\\
&&\hspace*{-4mm}=\hspace*{-0.5mm}\frac{\pi}{2}(-1)^{2j}(2j+1)\hspace*{-0.5mm}\Big{[}\Lambda^4+m_j^2\Lambda^2-m_j^4\text{log}\Big{(}\frac{2\Lambda}{m_j}\Big{)}+\mathcal{O}\Big{(}\frac{1}{\Lambda}\Big{)}\hspace*{-0.5mm}\Big{]},\nonumber
\end{eqnarray}
where $\Lambda$ is the momentum cutoff. Hence, for a compensation of the quartic, quadratic, and logarithmic, divergence one needs
\begin{eqnarray}
&\sum_j(-1)^{2j}(2j+1)=0,\quad\sum_j(-1)^{2j}(2j+1)m_j^2=0,&\nonumber\\
&\text{and}\,\,\,\sum_j(-1)^{2j}(2j+1)m_j^4=0.\,\,~&
\end{eqnarray}
This requires that the number of bosonic and fermionic degrees of freedom be equal, and the boson and fermion masses be degenerate.
Since these conditions seem to be met in the discrete supersymmetry model at zero temperature, the contributions from matter quantum fluctuations to the cosmological constant might indeed cancel in our construction.

We then conclude that discrete supersymmetry reduces, even in presence of potential energy contributions at the electroweak scale $M_\text{EW}\simeq 10^2\text{GeV}$,  the ratio between the expected and the observed vacuum energy density in the universe $\rho\simeq 10^{-47}\,(\text{GeV})^4$ \cite{Riess:1998cb,Perlmutter:1998np,Spergel:2003cb} by more than 60 orders of magnitude, since
\begin{equation}
(M_\text{Pl})^4/\rho\simeq 10^{119}\rightarrow (M_\text{EW})^4/\rho\simeq 10^{55}.\label{eq:reduction}
\end{equation}
Discrete supersymmetry therefore goes, in the exponent, half the way in scaling down the huge mismatch between the expected and the actual value of the cosmological constant.

Unbroken discrete supersymmetry may go even further by constraining the forms of the allowed scalar potentials and, thus, the possible contributions to the vacuum energy from potential energies. To see this, assume that at zero temperature the lowest energy of the Fermi-Bose system is given by a local minimum of the scalar potential and consider at some energy scale two local minima that preserve discrete supersymmetry. Now, as a consequence of the unbroken symmetry, the potential exhibits a dependence on the fluctuations around the values of the scalar field that is identical for both the expansions around the two minima. Therefore, if the potential is analytic and bounded from below, the potential energy differences between these minima will vanish, that is, the minima with unbroken discrete supersymmetry must be degenerate. This means that the contributions to the cosmological constant on the right-hand side in (\ref{eq:reduction}) coming from potential energy differences between the distinct supersymmetric minima are further reduced in the presence of discrete supersymmetry. Similar arguments would also apply to the contributions from fermion condensates that arise due to dynamical symmetry breaking during a deconfinement-confinement phase transition, such as in QCD, provided that discrete supersymmetry remains indeed unbroken at low energies.

\section{Conclusions}\label{sec:summary}

We have seen that a model with discrete supersymmetry yields a mechanism to have more matter and less vacuum energy and, thus, to address the combined dark matter and cosmological constant problems. One could have worried that the running of the visible and dark matter masses might introduce a breaking of the Fermi-Bose mass degeneracy established by discrete supersymmetry at the classical level, but this is not the case. At the quantum level, dangerous new interactions such as large flavor changing neutral currents, as in continuous supersymmetry \cite{Duncan:1983iq}, or Lorentz-violating couplings are absent. Instead, the model is in perfect agreement with the non-observation of superpartners at the LHC and other experiments \cite{Chatrchyan:2011zy,Aaltonen:2008rv,Aprile:2011hi,Hudson:2011zz}.

Discrete supersymmetry combines the advantages of continuous supersymmetry and gauge mirror models \cite{Foot:1995pa,Babu:2003is} or proposals involving a ghost sector \cite{Linde:1984ir,Kaplan:2005rr}, but differs from these also fundamentally. First, there have been several intriguing suggestions to solve the cosmological constant problem using unbroken continuous supersymmetry \cite{Witten:1994cga,Dvali:2000ty}, but these usually rely on other than four dimensions. Next, while mirror models can include a discrete exchange symmetry between the standard model and the mirror sector, this kind of symmetry can only connect particles that have the same spin and norm, and would not guarantee a cancellation between the vacuum energies coming from the mirror universes. In scenarios with a ghost sector, on the other hand, a discrete symmetry, ``energy parity'', between usual and ghost fields could account for the cosmological constant problem, but seems to require a low-energy modification of gravity to prohibit rapid vacuum decay \cite{Carroll:2003st,Cline:2003gs}. Discrete supersymmetry transformations that may yield a small cosmological constant have also been proposed in string theory and brane models, but these are either limited to a graviton-dilatino-sector and do not apply to the full low-energy theory \cite{Zwirner:1988fn} or demand solely gravitational interactions between the usual fields \cite{Burgess:1999qw} .

Different from the continuous case, a Grassmann-valued supersymmetry parameter is absent. Our model shares this aspect with continuous unitary supersymmetry \cite{Cahill:2001tt}, but avoids leading from purely bosonic or fermionic states to a superposition of states of even and odd fermion numbers, which are forbidden by a superselection rule. When comparing, on the other hand, with the possible continuous supercharges as given by the Haag-Lopuszanski-Sohnius theorem \cite{Haag:1974qh}, one has to bear in mind that the fermionic symmetry generators ``carry a representation'' \cite{Sohnius:1985qm} of the bosonic generators. The latter are, in continuous supersymmetry, fixed by the Coleman-Mandula theorem \cite{Coleman:1967ad}, which deals with infinitesimal symmetries that are described by a Lie algebra (for a detailed discussion see \cite{Weinberg:2000cr}) and is not applicable to discrete symmetries. Since in our case the generators are discrete, the conventional description and classification of the supercharges does, therefore, not apply to discrete supersymmetry.

The discrete supersymmetry transformation is linear, independent from any parameter in the Lagrangian and transforms the kinetic, mass, and interaction terms independently. In fact, discrete supersymmetry possesses an off-shell formulation, which facilitates the application of supersymmetric Ward identities and is necessary for a manifestly supersymmetric quantization \cite{Gates:1983nr}. Actually, in our model, we have applied discrete supersymmetry to simplify loop calculations in the dark matter sector.

It can be shown that unbroken discrete supersymmetry ensures at zero temperature that the potential energy differences between supersymmetric minima of the scalar potential vanish. On the other hand, an investigation of how the implications of supersymmetric QCD \cite{Intriligator:1995au} for the vacuum energy can be represented in the discrete model would be called for in some future work. Also, in order to explain the non-zero vacuum energy density in the universe of the order $10^{-47}\,(\text{GeV})^4$, one may invoke additional ideas \cite{Wetterich:1987fm}. We nevertheless share the position that the true problem is not to understand the actual nonzero value of the cosmological constant, but why it is not at least 60 orders of magnitude larger \cite{Peskin:2011zz}.

It is clear that our mechanism can be applied in the same fashion to the actual standard model fields and their interactions without difficulty, and supersymmetry guarantees that the feature of canceling vacuum energies is still intact. The same applies to extensions of the standard model, such as unified field theories \cite{Georgi:1974sy}. One may now ask how to combine the advantages of discrete supersymmetry with continuous supersymmetry. We have nothing new to say about continuous supersymmetry; one could, thus, use the minimal supersymmetric standard model \cite{Dimopoulos:1981zb}, or variants thereof, with a discrete supersymmetry copy to cancel vacuum energies. This allows also to ensure enhanced gauge coupling unification \cite{Amaldi:1991cn} and have extra sources for dark matter. Notwithstanding, considering the recent discovery of the Higgs boson, the gauge hierarchy problem could in our construction also be addressed by alternative mass generation mechanisms \cite{Nambu:1961tp,alternativemass}, where the Higgs is a composite field \cite{Kaplan:1983fs,Giudice:2007fh,Marzocca:2012zn}.

\section*{Acknowledgements}
I would like to thank K.S.~Babu, C.~Csaki, M. Drees, and F.P.~Schuller for useful comments, discussions, and suggestions on the text.

\appendix


\begin{thebibliography}{00}

\bibitem{Ade:2013ktc} 
  PLANCK Collaboration, P.~A.~R.~Ade {\it et al.},
  Astron.\ Astrophys.\  {\bf 571}, A1 (2014); Astron.\ Astrophys.\  {\bf 571}, A16 (2014).
 
 \bibitem{Riess:1998cb}
  Supernova Search Team Collaboration, A.~G.~Riess {\it et al.},
  Astron.\ J.\  {\bf 116}, 1009-1038 (1998);  J.~L.~Tonry {\it et al.},
  Astrophys.\ J.\  {\bf 594}, 1-24 (2003).
  
\bibitem{Perlmutter:1998np}
  Supernova Cosmology Project Collaboration,
  S.~Perlmutter {\it et al.},
  Astrophys.\ J.\  {\bf 517}, 565-586 (1999); R.~A.~Knop {\it et al.},
  Astrophys.\ J.\  {\bf 598}, 102 (2003).  
  
\bibitem{Spergel:2003cb}
  WMAP Collaboration, D.~N.~Spergel {\it et al.},
  Astrophys.\ J.\ Suppl.\  {\bf 148}, 175-194 (2003).
  
\bibitem{Weinberg:1988cp}
  S.~Weinberg,
  Rev.\ Mod.\ Phys.\  {\bf 61}, 1 (1989); arXiv:astro-ph/0005265.
  
\bibitem{Carroll:2000fy} 
  S.~M.~Carroll,
  Living Rev.\ Rel.\  {\bf 4}, 1 (2001).
  
\bibitem{Nobbenhuis:2004wn}
  S.~Nobbenhuis,
  Found.\ Phys.\  {\bf 36}, 613 (2006).
 
\bibitem{Nicolai:1976xp} 
  H.~Nicolai,
  J.\ Phys.\ A {\bf 9}, 1497 (1976).
  
 \bibitem{Witten:1981nf}
  E.~Witten,
  Nucl.\ Phys.\  B {\bf 188}, 513 (1981).
  
\bibitem{Gendenshtein:1986ub} 
  L.~E.~Gendenshtein and I.~V.~Krive,
  Sov.\ Phys.\ Usp.\  {\bf 28}, 645 (1985)
  [Usp.\ Fiz.\ Nauk {\bf 146}, 553 (1985)].
  
\bibitem{Wess:1974tw}
  J.~Wess and B.~Zumino,
  Nucl.\ Phys.\  B {\bf 70}, 39 (1974).

\bibitem{Oldham:1974} 
  K.~B.~Oldham and J.~Spanier, {\it The Fractional Calculus},
 (Academic Press, New York,  1974);
  S.~G.~Samko, A.~A.~Kilbas, and O.~I.~Marichev, {\it Fractional Integrals and Derivatives},
 (Gordon and Breach, Yverdon, 1993).
 
\bibitem{Derksen:2005}
H.~Derksen and J.~Weyman, AMS Notices {\bf 52}, 200 (2005).
 
\bibitem{Gates:1983nr} 
 S.~J.~Gates, M.~T.~Grisaru, M.~Rocek and W.~Siegel,
 Front.\ Phys.\  {\bf 58}, 1 (1983).
 
\bibitem{Sohnius:1985qm}
  M.~F.~Sohnius,
  Phys.\ Rept.\  {\bf 128}, 39 (1985).   
 
 \bibitem{Wigner:1931}
 E.~P.~Wigner, {\it Gruppentheorie und ihre Anwendung auf die Quantenmechanik der Atomspektren} (Vieweg, Braunschweig, 1931): pp.~251--253.
 
\bibitem{Peskin:1995ev} 
  M.~E.~Peskin and D.~V.~Schroeder,
  {\it An Introduction to Quantum Field Theory}
 (Addison-Wesley, Reading, 1995): Chapter 3.5.
 
\bibitem{Weinberg:1995mt} 
  S.~Weinberg,
  {\it The Quantum Theory of Fields. Vol. 1: Foundations}
  (Cambridge University Press, Cambridge, 1995): Chapter 5.1.
 
\bibitem{Haag:1955ev} 
  R.~Haag,
  Kong.\ Dan.\ Vid.\ Sel.\ Mat.\ Fys.\ Med.\  {\bf 29N12}, 1 (1955).

\bibitem{Sakurai:2011}
J.~J.~Sakurai and J.~J.~Napolitano, {\it Modern Quantum Mechanics} (Addison-Wesley, San Francisco, 2011): Chapters 2.1 and 5.7.

\bibitem{Soroka:1995et} 
  V.~A.~Soroka,
  Phys.\ Atom.\ Nucl.\  {\bf 59}, 1270 (1996)
  [Yad.\ Fiz.\  {\bf 59}, 1327 (1996)];  D.~V.~Soroka, V.~A.~Soroka and J.~Wess,
  Phys.\ Lett.\ B {\bf 512}, 197 (2001);
  R.~Casalbuoni, F.~Elmetti, S.~Knapen and L.~Tamassia,
  JHEP {\bf 1004}, 106 (2010).
 
\bibitem{Nambu:1961tp} 
 Y.~Nambu and G.~Jona-Lasinio, Phys.\ Rev.\  {\bf 122}, 345 (1961). 
 
 \bibitem{ArkaniHamed:1998rs}
  N.~Arkani-Hamed, S.~Dimopoulos and G.~R.~Dvali,
  Phys.\ Lett.\  B {\bf 429}, 263 (1998);  Phys.\ Rev.\  D {\bf 59}, 086004 (1999);
  I.~Antoniadis, N.~Arkani-Hamed, S.~Dimopoulos and G.~R.~Dvali,
  Phys.\ Lett.\  B {\bf 436}, 257 (1998).
  
\bibitem{Randall:1999ee}
  L.~Randall and R.~Sundrum,
  Phys.\ Rev.\ Lett.\  {\bf 83}, 3370 (1999);
  Phys.\ Rev.\ Lett.\  {\bf 83}, 3370 (1999).    

\bibitem{Haber:1984rc}
  H.~E.~Haber and G.~L.~Kane,
  Phys.\ Rept.\  {\bf 117}, 75 (1985).
  
  \bibitem{Denner:1992vza}
  A.~Denner, H.~Eck, O.~Hahn and J.~Kublbeck,
  Nucl.\ Phys.\  B {\bf 387}, 467 (1992);
  Phys.\ Lett.\  B {\bf 291}, 278 (1992).
  
\bibitem{Mohapatra:1998rq}
  R.~N.~Mohapatra and P.~B.~Pal,
 {\it Massive Neutrinos in Physics and Astrophysics}
 (World Scientific, Singapore, 2004): pp. 73--74.
 
\bibitem{Dreiner:2008tw} 
  H.~K.~Dreiner, H.~E.~Haber and S.~P.~Martin,
  Phys.\ Rept.\  {\bf 494}, 1 (2010).
  
\bibitem{Froggatt:1978nt} 
  C.~D.~Froggatt and H.~B.~Nielsen,
  Nucl.\ Phys.\ B {\bf 147}, 277 (1979).

\bibitem{Beringer:1900zz} 
  Particle Data Group Collaboration, J.~Beringer {\it et al.}, Phys.\ Rev.\ D {\bf 86}, 010001 (2012)
  and 2013 update for the 2014 edition (URL: http://pdg.lbl.gov).

\bibitem{Peskin:1995ev2} 
 Peskin, {\it Quantum Field Theory}: Chapter 12.4. 

\bibitem{Krauss:1988zc} 
  L.~M.~Krauss and F.~Wilczek,
  Phys.\ Rev.\ Lett.\  {\bf 62}, 1221 (1989).

\bibitem{Grisaru:1979wc} 
  M.~T.~Grisaru, W.~Siegel and M.~Rocek,
  Nucl.\ Phys.\ B {\bf 159}, 429 (1979).
  
\bibitem{Bernabei:2003za}
  R.~Bernabei {\it et al.},
  Riv.\ Nuovo Cim.\  {\bf 26N1}, 1 (2003);
  DAMA Collaboration, R.~Bernabei {\it et al.},
  Eur.\ Phys.\ J.\  C {\bf 56}, 333 (2008).
  
\bibitem{Ahmed:2009zw}
 CDMS-II Collaboration,
  Z.~Ahmed {\it et al.},
  Science {\bf 327}, 1619 (2010).

\bibitem{Angle:2007uj}
XENON Collaboration, J.~Angle {\it et al.},
  Phys.\ Rev.\ Lett.\  {\bf 100}, 021303 (2008);
  XENON10 Collaboration,
  J.~Angle {\it et al.},
  Phys.\ Rev.\  D {\bf 80}, 115005 (2009).
  
\bibitem{Trimble:1987ee}
  V.~Trimble,
  Ann.\ Rev.\ Astron.\ Astrophys.\  {\bf 25}, 425 (1987).

\bibitem{Hooper:2009zm}
  D.~Hooper, arXiv:0901.4090.

\bibitem{Boehm:2003hm} 
  C.~Boehm and P.~Fayet,
  Nucl.\ Phys.\ B {\bf 683}, 219 (2004);
  M.~Casse and P.~Fayet,
  astro-ph/0510490.
  
\bibitem{Pospelov:2008jk} 
  M.~Pospelov, A.~Ritz and M.~B.~Voloshin,
  Phys.\ Rev.\ D {\bf 78}, 115012 (2008);
  Phys.\ Lett.\ B {\bf 662}, 53 (2008).
  
\bibitem{Foot:2003iv} 
  R.~Foot, Phys.\ Rev.\ D {\bf 69}, 036001 (2004);  arXiv:1203.2387 [hep-ph].
  
\bibitem{Lee:1977ua} 
  B.~W.~Lee and S.~Weinberg,
  Phys.\ Rev.\ Lett.\  {\bf 39}, 165 (1977).
  
\bibitem{Holdom:1985ag} 
  B.~Holdom,
  Phys.\ Lett.\ B {\bf 166}, 196 (1986).
  
\bibitem{ArkaniHamed:2008qn} 
  N.~Arkani-Hamed, D.~P.~Finkbeiner, T.~R.~Slatyer and N.~Weiner,
  Phys.\ Rev.\ D {\bf 79}, 015014 (2009); 
  R.~Essig, J.~A.~Jaros, W.~Wester, P.~H.~Adrian, S.~Andreas, T.~Averett, O.~Baker and B.~Batell {\it et al.},
  arXiv:1311.0029 [hep-ph].
  
\bibitem{Witten:1982df} 
  E.~Witten,
  Nucl.\ Phys.\ B {\bf 202}, 253 (1982).
  
\bibitem{Cohen:1988kt} 
  A.~G.~Cohen and D.~B.~Kaplan,
  Nucl.\ Phys.\ B {\bf 308}, 913 (1988).
     
\bibitem{Zumino:1974bg} 
  B.~Zumino,
  Nucl.\ Phys.\ B {\bf 89}, 535 (1975).
  
\bibitem{Zeldovich:1967gd} 
  Y.~B.~Zeldovich,
  JETP Lett.\  {\bf 6}, 316 (1967);
  Sov.\ Phys.\ Usp.\  {\bf 11}, 381 (1968).
  
\bibitem{Ramond:1971gb} 
  P.~Ramond, Phys.\ Rev.\ D {\bf 3}, 2415 (1971).
  
 \bibitem{Golfand:1971iw}
  Yu.~A.~Golfand and E.~P.~Likhtman,
  JETP Lett.\  {\bf 13}, 323 (1971)
  [Pisma Zh.\ Eksp.\ Teor.\ Fiz.\  {\bf 13}, 452 (1971)].
   
\bibitem{Volkov:1972jx}
  D.~V.~Volkov and V.~P.~Akulov,
  JETP Lett.\  {\bf 16}, 438 (1972)
  [Pisma Zh.\ Eksp.\ Teor.\ Fiz.\  {\bf 16}, 621 (1972)].

\bibitem{Nilles:1983ge}
  H.~P.~Nilles,
  Phys.\ Rept.\  {\bf 110}, 1 (1984).
  
\bibitem{Wess:1992cp} 
  J.~Wess and J.~Bagger, {\it Supersymmetry and Supergravity}
  (Princeton University Press, Princeton, 1992).  
  
\bibitem{Martin:1997ns}
  S.~P.~Martin,
  arXiv:hep-ph/9709356.  

\bibitem{Pauli:2000}
 W.~Pauli, {\it  Pauli Lectures on Physics Vol. 6: Selected Topics in Field Quantization}
(Dover Publications, New York, 2000): Section 9.

\bibitem{West:1986wua}
  P.~C.~West, {\it Introduction to Supersymmetry and Supergravity}
(World Scientific, Singapore, 1986): Chapter 3.

\bibitem{Duncan:1983iq} 
  M.~J.~Duncan,
  Nucl.\ Phys.\ B {\bf 221}, 285 (1983);
  J.~F.~Donoghue, H.~P.~Nilles and D.~Wyler,
  Phys.\ Lett.\ B {\bf 128}, 55 (1983); 
  A.~Bouquet, J.~Kaplan and C.~A.~Savoy,
  Phys.\ Lett.\ B {\bf 148}, 69 (1984).
  
\bibitem{Chatrchyan:2011zy} 
CMS Collaboration,  S.~Chatrchyan {\it et al.},
 Phys.\ Rev.\ Lett.\  {\bf 107}, 221804 (2011);
 ATLAS Collaboration, G.~Aad {\it et al.},
Phys.\ Lett.\ B {\bf 709}, 137 (2012).
      
\bibitem{Aaltonen:2008rv}
CDF Collaboration, T.~Aaltonen {\it et al.},
  Phys.\ Rev.\ Lett.\  {\bf 102}, 121801 (2009);
 D0 Collaboration, V.~M.~Abazov {\it et al.},
  Phys.\ Lett.\  B {\bf 660}, 449 (2008);
  Phys.\ Lett.\  B {\bf 680}, 34 (2009).
  
\bibitem{Aprile:2011hi}
  XENON100 Collaboration, E.~Aprile {\it et al.},
  Phys.\ Rev.\ Lett.\  {\bf 107}, 131302 (2011).   

\bibitem{Hudson:2011zz}
  J.~J.~Hudson, D.~M.~Kara, I.~J.~Smallman, B.~E.~Sauer, M.~R.~Tarbutt, E.~A.~Hinds,
  Nature {\bf 473}, 493-496 (2011).
  
\bibitem{Foot:1995pa}
R.~Foot and R.R.~Volkas,
Phys. Rev. D {\bf 52} (1995) 6595;
Z.G. Berezhiani and R.N. Mohapatra,
Phys. Rev. D {\bf 52} (1995) 6607.

\bibitem{Babu:2003is}
  K.~S.~Babu and G.~Seidl,
  Phys.\ Lett.\  B {\bf 591}, 127 (2004);
  Phys.\ Rev.\  D {\bf 70}, 113014 (2004).
  
\bibitem{Linde:1984ir} 
  A.~D.~Linde,
  Rept.\ Prog.\ Phys.\  {\bf 47}, 925 (1984).
  
\bibitem{Kaplan:2005rr} 
  D.~E.~Kaplan and R.~Sundrum,
  JHEP {\bf 0607}, 042 (2006).  
  
\bibitem{Witten:1994cga}
  E.~Witten,
  Int.\ J.\ Mod.\ Phys.\  A {\bf 10}, 1247 (1995);  arXiv:hep-ph/0002297.
 
 \bibitem{Dvali:2000ty}
  G.~R.~Dvali, arXiv:hep-th/0004057.
  
\bibitem{Carroll:2003st} 
  S.~M.~Carroll, M.~Hoffman and M.~Trodden,
  Phys.\ Rev.\ D {\bf 68}, 023509 (2003).
  
\bibitem{Cline:2003gs} 
  J.~M.~Cline, S.~Jeon and G.~D.~Moore,
  Phys.\ Rev.\ D {\bf 70}, 043543 (2004).
    
 \bibitem{Zwirner:1988fn} 
  F.~Zwirner, Talk given at the {\it 23rd Rencontres de Moriond Conference - Electroweak Interactions and Unified Theories}, Les Arcs, France, 6-13 March 1988, UCB-PTH-88-12, LBL-25278, C88-03-06.1.  

\bibitem{Burgess:1999qw} 
  C.~P.~Burgess, R.~C.~Myers and F.~Quevedo,
  Phys.\ Lett.\ B {\bf 495}, 384 (2000). 

\bibitem{Cahill:2001tt} 
  K.~E.~Cahill,
  JHEP {\bf 0106}, 002 (2001).
  
\bibitem{Haag:1974qh}
  R.~Haag, J.~T.~Lopuszanski and M.~Sohnius,
  Nucl.\ Phys.\  B {\bf 88}, 257 (1975).   
  
\bibitem{Coleman:1967ad}
  S.~R.~Coleman and J.~Mandula,
  Phys.\ Rev.\  {\bf 159}, 1251 (1967).

\bibitem{Weinberg:2000cr} 
  S.~Weinberg, {\it The Quantum Theory of Fields. Vol. 3: Supersymmetry}
  (Cambridge University Press, Cambridge, 2000): Chapter 24.
  
\bibitem{Intriligator:1995au} 
  K.~A.~Intriligator and N.~Seiberg,
  Nucl.\ Phys.\ Proc.\ Suppl.\  {\bf 45BC}, 1 (1996).

\bibitem{Wetterich:1987fm}
  C.~Wetterich,
  Nucl.\ Phys.\  B {\bf 302}, 668 (1988);  B.~Ratra and P.~J.~E.~Peebles,
  Phys.\ Rev.\  D {\bf 37}, 3406 (1988).

\bibitem{Peskin:2011zz} 
  M.~E.~Peskin,
  Pramana {\bf 79}, 1169 (2012).

\bibitem{Georgi:1974sy} 
  H.~Georgi and S.~L.~Glashow,
  Phys.\ Rev.\ Lett.\  {\bf 32}, 438 (1974);
  J.~C.~Pati and A.~Salam,
  Phys.\ Rev.\ D {\bf 10}, 275 (1974)
  [Erratum-ibid.\ D {\bf 11}, 703 (1975)].
  
\bibitem{Dimopoulos:1981zb}
  S.~Dimopoulos and H.~Georgi,
  Nucl.\ Phys.\  B {\bf 193}, 150 (1981);
  S.~Dimopoulos, S.~Raby and F.~Wilczek,
  Phys.\ Rev.\  D {\bf 24}, 1681 (1981).
  
\bibitem{Amaldi:1991cn} 
  U.~Amaldi, W.~de Boer and H.~Furstenau,
  Phys.\ Lett.\ B {\bf 260}, 447 (1991).

\bibitem{alternativemass} 
S. Weinberg, Phys.\ Rev.\ D {\bf 13}, 974 (1976): L. Susskind, Phys.\ Rev.\ D {20}, 2619 (1979); W. A. Bardeen, C. T. Hill and M. Lindner, Phys.\ Rev.\ D {\bf 41}, 1647 (1990); C. Csaki, C. Grojean, H. Murayama, L. Pilo and J. Terning, Phys.\ Rev.\ D {\bf 69}, 055006 (2004).
  
\bibitem{Kaplan:1983fs} 
  D.~B.~Kaplan and H.~Georgi,
  Phys.\ Lett.\ B {\bf 136}, 183 (1984); M.~J.~Dugan, H.~Georgi and D.~B.~Kaplan,
  Nucl.\ Phys.\ B {\bf 254}, 299 (1985).

\bibitem{Giudice:2007fh} 
  G.~F.~Giudice, C.~Grojean, A.~Pomarol and R.~Rattazzi,
 JHEP {\bf 0706}, 045 (2007).

\bibitem{Marzocca:2012zn} 
  D.~Marzocca, M.~Serone and J.~Shu,
  JHEP {\bf 1208}, 013 (2012);
  R.~Barbieri, D.~Buttazzo, F.~Sala, D.~M.~Straub and A.~Tesi,
  JHEP {\bf 1305}, 069 (2013).
    
\end{thebibliography}
\end{document}